\documentclass[11pt]{article}
\usepackage{graphicx}
\usepackage{amsmath}
\usepackage{amssymb}
\usepackage{amsxtra}

\def\institute#1{\gdef\@institute{#1}}
\let\oldmaketitle\maketitle
\renewcommand\maketitle{\oldmaketitle\noindent\@institute}

\begin{document}
\allowdisplaybreaks

\title{Exact expressions for the renormalization constants in the MS-like schemes}
\author{K.V.~Stepanyantz$^{1}$}
\institute{
$^1$  Moscow State University, 119991, Moscow, Russia
}

\maketitle

\begin{abstract}
We briefly review how it is possible to derive some exact expressions for the renormalization constants for the MS-like renormalization prescriptions using the arguments based on the renormalization group. These expressions are obtained for a version of the dimensional technique in which the dimensionful parameter $\Lambda$ differs from the renormalization scale $\mu$. They encode the equations relating the coefficients at higher $\varepsilon$-poles, powers of $\ln \Lambda/\mu$, and mixed terms of the structure $\varepsilon^{-q} \ln^p \Lambda/\mu$ to the coefficients of the renormalization group functions (i.e. of the $\beta$-function and the anomalous dimension). The general results are verified by some multiloop calculations.
\end{abstract}

\section{Introduction}

Quantum corrections are very important for understanding nature. For example, a very precise agreement of the electron anomalous magnetic moment value with the theoretical prediction is one of the most convincing arguments in favour of the fact that nature is described by quantum field theory \cite{Peskin:1995ev}. The unification of running couplings is an indirect evidence in favour of supersymmetry and Grand Unification \cite{Mohapatra:1986uf} obtained by combining the experimental values of the coupling constants and theoretical calculations of quantum corrections \cite{Ellis:1990wk,Amaldi:1991cn,Langacker:1991an}.

In most quantum field theory models quantum corrections diverge in the ultraviolet (UV) region, so that a theory for which they are investigated should be properly regularized \cite{Gnendiger:2017pys}. The most popular method for this purpose is dimensional regularization \cite{'tHooft:1972fi,Bollini:1972ui,Ashmore:1972uj,Cicuta:1972jf} when a theory is considered in the non-integer dimension $D\equiv 4-\varepsilon$. In this case divergences appear as $\varepsilon$-poles. These $\varepsilon$-poles should removed by the renormalization of some parameters, e.g., couplings, masses, fields, etc. This is possible in the renormalizable theories.

However, in the supersymmetric case the dimensional regularization is not convenient because supersymmetry is explicitly broken \cite{Jack:1997sr}. That is why for supersymmetric theories it is more convenient to use a modification of this technique called dimensional reduction \cite{Siegel:1979wq}. Nevertheless, sometimes it is better to use the regulations of the cutoff type, e.g., the higher covariant derivative regularization \cite{Slavnov:1971aw,Slavnov:1972sq,Slavnov:1977zf}.  In this case (logarithmic) divergences are given by powers of $\ln\Lambda/\mu$, where $\Lambda$ is a dimensionful regularization parameter, and $\mu$ is the renormalization point. The superfield formulation of the higher covariant derivative method \cite{Krivoshchekov:1978xg,West:1985jx} is especially useful for supersymmetric theories, because
it does not break supersymmetry and reveals some interesting features of quantum corrections  \cite{Stepanyantz:2019lyo}. For instance, the exact Novikov, Shifman, Vainshtein, and Zakharov (NSVZ) $\beta$-function \cite{Novikov:1983uc,Jones:1983ip,Novikov:1985rd,Shifman:1986zi} is naturally obtained with this regularization. According to \cite{Stepanyantz:2016gtk,Stepanyantz:2019ihw,Stepanyantz:2020uke}, the NSVZ $\beta$-function is valid in all loops if a supersymmetric theory is regularized by Higher covariant Derivatives and renormalization is made by Minimal Subtraction of Logarithms (the so-called HD+MSL scheme), see \cite{Stepanyantz:2017sqg} and references therein. In the Abelian case this result was derived earlier in \cite{Stepanyantz:2011jy,Kataev:2013eta}. Note that in the $\overline{\mbox{DR}}$-scheme the NSVZ equation is not valid starting from the approximation where the scheme dependence becomes essential \cite{Jack:1996vg,Jack:1996cn,Jack:1998uj}. However, in this case the NSVZ relation can be restored in each order of the perturbation theory with the help of a finite renormalization of the coupling constant, because some scheme-independent consequences of the NSVZ equation are satisfied \cite{Kataev:2013csa,Kataev:2014gxa}. This implies that the NSVZ equation is valid only for a special class of renormalization prescriptions \cite{Goriachuk:2018cac}, some of them being naturally constructed with the help of the higher covariant derivative regularization.

Note that the $\varepsilon$-poles in the renormalization constants obtained with the help of the dimensional technique are analogs of $\ln\Lambda/\mu$ in the corresponding expressions for the renormalization constants derived using the regularizations of the cutoff type. The relation between the simple poles and the first power of this logarithm has a very simple form, e.g.,

\begin{equation}\label{Simple_Poles}
\ln Z_\alpha = \smash{-\sum\limits_{L=1}^\infty \alpha^L \beta_L \Big(\frac{1}{L\varepsilon} + \ln\frac{\Lambda}{\mu}\Big) + \mbox{higher poles and logarithms}},
\end{equation}

\noindent
where $L$ is a number of loops \cite{Chetyrkin:1980sa}. However, the coefficients at higher poles and logarithms are related in a rather nontrivial way and no simple relations are seen at the first sight. In this paper the corresponding relations will be constructed in all orders. For this purpose we will use the arguments based on the structure of a group formed by finite renormalizations.

\section{The group formed by finite renormalizations}
\label{Section_Group_Structure}

In renormalizable quantum field theory models UV divergences can be removed by the renormalization, e.g.,

\begin{equation}
\alpha_0 = \alpha_0(\alpha(\mu),\ln\Lambda/\mu);\qquad \varphi = Z(\alpha,\ln\Lambda/\mu) \varphi_{R},
\end{equation}

\noindent
where $\mu$ is a renormalization point and $\Lambda$ is the dimensionful parameter introduced by a regularization. However, the way of making the renormalization is not uniquely defined. In general, various renormalization prescriptions are related by the finite renormalizations  \cite{Kallen:1954jfh,Vladimirov:1975mx,Vladimirov:1979ib} of the form

\begin{equation}\label{Finite_Renormalization_General_Form}
\alpha' = \alpha'(\alpha);\qquad Z'(\alpha',\ln\Lambda/\mu) = z(\alpha) Z(\alpha,\ln\Lambda/\mu).
\end{equation}

\noindent
Note that the renormalization group functions (RGFs) nontrivially transform under these finite renormalizations \cite{Vladimirov:1979my}. Namely, two first coefficients of the gauge $\beta$-function and the first coefficient of the anomalous dimension are scheme independent, while the other coefficients depend on a specific choice of the renormalization prescription.

It is easy to see that finite renormalizations (\ref{Finite_Renormalization_General_Form}) form an infinite dimensional Lie group. Following \cite{Kataev:2024xbl}, to describe it, we first construct its algebra using the exponential map \cite{Isaev:2018xcg}. For this purpose we recall that if ${\cal G}$ is a Lie group, then in a certain vicinity of the identity element 1 it is possible to present the group element $\hat\omega \in {\cal G}$ as  the exponential of the corresponding Lie algebra ${\cal A}$ element $\hat a \in {\cal A}$, $\hat\omega = \exp(\hat a)$. Therefore, it is expedient to consider an infinitesimal finite renormalization of the coupling constant $\alpha\to \alpha'(\alpha)$ and present it as the series

\begin{equation}\label{Alpha_Tranformations_Small}
\delta\alpha = -\sum\limits_{n=1}^\infty a_n \alpha^{n+1} \equiv \sum\limits_{n=1}^\infty a_n \hat L_n \alpha \equiv \hat a \alpha,
\end{equation}

\noindent
where $a_n$ are arbitrary real parameters. The operators $\hat L_n$ with $n\ge 1$ are the generators of the group of finite renormalizations. In the explicit form these generators (acting on an arbitrary function of $\alpha$) are written as

\begin{equation}
\hat L_n = - \alpha^{n+1} \frac{d}{d\alpha}
\end{equation}

\noindent
and satisfy the commutation relations of the Witt algebra\footnote{The relation between the renormalization group and the Witt algebra was also discussed in \cite{Isaev:2018xcg}.}

\begin{equation}
[\hat L_n,\hat L_m] = (n-m) \hat L_{n+m}.
\end{equation}

\noindent
However, in the Witt algebra $n$ is an arbitrary integer, while in the case under consideration $n\ge 1$ due to the use of the perturbation theory. Therefore, the Lie algebra of the group formed by finite renormalizations (for the renormalization of charge) is a subalgebra of the Witt algebra spanned by $\hat L_n$ with $n\ge 1$.

The exponential map allows obtaining the non-infinitesimal transformations,

\begin{equation}
\alpha' =\hat\omega\alpha = \exp\Big(\sum\limits_{n=1}^\infty a_n \hat L_n\Big) \alpha = \exp\Big(-\sum\limits_{n=1}^\infty a_n \alpha^{n+1} \frac{d}{d\alpha}\Big) \alpha.
\end{equation}

\noindent
The explicit form of these transformations in the lowest approximations can be found in \cite{Kataev:2024xbl}.

The finite renormalizations that include the renormalization of the matter fields and masses can be considered similarly. According to (\ref{Finite_Renormalization_General_Form}), they are determined by the functions $\alpha'(\alpha)$ and $z(\alpha)$. For the infinitesimal finite renormalizations $z(\alpha)$ can be presented in the form

\begin{equation}
z(\alpha) = 1 - \sum\limits_{n=1}^\infty z_n \alpha^n + O(az,z^2).
\end{equation}

\noindent
This implies that the renormalized fields change as

\begin{equation}
\varphi_R' = z^{-1}(\alpha)\varphi_R = \Big(1 + \smash{\sum\limits_{n=1}^\infty} z_n \alpha^n + O(z^2)\Big)\varphi_R \equiv \varphi_R + \smash{\sum\limits_{n=1}^\infty} z_n \hat G_n \varphi_R + O(z^2),
\end{equation}

\noindent
where we have introduced the operators

\begin{equation}
\hat G_n \varphi_R \equiv \alpha^n \varphi_R,
\end{equation}

\noindent
which satisfy the commutation relations

\begin{equation}
[\hat L_n,\hat L_m] = (n-m) \hat L_{n+m};\quad [\hat G_n,\hat G_m] = 0;\quad [\hat L_n,\hat G_m] = - m\hat G_{n+m},
\end{equation}

\noindent
where $n,m\ge 1$.

The non-infinitesimal transformations can again be constructed with the help of the exponential map,

\begin{equation}
z(\alpha) = \exp\Big(\sum\limits_{n=1}^\infty a_n \hat L_n\Big) \exp\Big(-\sum\limits_{n=1}^\infty a_n \hat L_n - \sum\limits_{n=1}^\infty z_n \hat G_n\Big).
\end{equation}

\section{Rescaling subgroup}
\label{Section_Rescaling}

An important particular case of the finite renormalizations is the transformations changing the renormalization scale.

\begin{eqnarray}
&& \alpha(\mu) \to \alpha(\mu')\equiv \alpha'(\mu);\qquad\nonumber\\
&& Z\big(\alpha(\mu),\ln\Lambda/\mu\big) \to Z'\big(\alpha(\mu'),\ln\Lambda/\mu'\big) \equiv z\big(\alpha(\mu)\big) Z\big(\alpha(\mu),\ln\Lambda/\mu\big)\qquad
\end{eqnarray}

\noindent
parameterized by $t = \ln \mu'/\mu$. Evidently, these transformations form an Abelian group, which is a subgroup of the group composed of finite renormalizations. For the infinitesimal transformations for which $\mu'$ is close to $\mu$ (or, equivalently, the parameter $t$ is small)

\begin{eqnarray}
&& \alpha'-\alpha = \beta(\alpha) t + O(t^2)\nonumber\\
&&\qquad\qquad = t \sum\limits_{n=1}^\infty \beta_n \alpha^{n+1} + O(t^2)
= - t \sum\limits_{n=1}^\infty \beta_n \hat L_n\alpha + O(t^2);\nonumber\\
&& \varphi_R'-\varphi_R = - \delta z \varphi_R + O(t^2) = -\gamma(\alpha) t\, \varphi_R + O(t^2)\nonumber\\
&&\qquad\, = - t\sum\limits_{n=1}^\infty \gamma_n \alpha^n\, \varphi_R + O(t^2) = - t \sum\limits_{n=1}^\infty \gamma_n \hat G_n \varphi_R + O(t^2),\qquad
\end{eqnarray}

\noindent
so that the generator of the rescaling transformations is the operator

\begin{equation}
\hat L = - \sum\limits_{n=1}^\infty \Big(\beta_n \hat L_n + \gamma_n \hat G_n\Big).
\end{equation}

Applying the exponential map for constructing the non-infinitesimal transformations of the coupling constant we obtain that an arbitrary function of the coupling constant changes as \cite{Groote:2001im,Mikhailov:2004iq}.

\begin{equation}
f(\alpha') = \exp\Big(\ln\frac{\mu'}{\mu}\, \beta(\alpha)\frac{\partial}{\partial\alpha}\Big) f(\alpha).
\end{equation}

\noindent
However, it is easy to see that under the rescaling transformations RGFs remains unchanged.

\section{Exact expresssions for the renormalization constants}
\label{Section_Renormalization_Constants}

As well known, in various renormalization constants the coefficients at higher $\varepsilon$-poles are related to the coefficients of RGFs \cite{tHooft:1973mfk} by the 't Hooft pole equations, see \cite{Kazakov:2008tr} for  review. This is also valid for the coefficients at higher powers of logarithms \cite{Collins:1984xc}. The argumentation based on the algebraic structure of the rescaling subgroup allows to construct simple and beautiful expressions for renormalization constants that relate them to the renormalization group functions.

It is especially interesting to consider such a regularization when both $\varepsilon$-poles and logarithms are present in the renormalization constants. This structure of divergent contributions can be obtained for a special modification of dimensional regularization. In the dimension $D\ne 4$ the bare gauge coupling constant $\widetilde\alpha_0$ has the dimension $m^\varepsilon$ and can, therefore, be presented as $\widetilde\alpha_0 = \alpha_0\Lambda^\varepsilon$, where $\alpha_0$ is dimensionless and $\Lambda$ is a parameter with the dimension of mass. Then the coupling constant can be renormalized according to the prescription

\begin{equation}
\alpha_0 = \Big(\frac{\mu}{\Lambda}\Big)^{\varepsilon} \mbox{\boldmath$\alpha$}\, \mbox{\boldmath$Z_\alpha$}^{-1}(\mbox{\boldmath$\alpha$}, \varepsilon^{-1}),
\end{equation}

\noindent
where $\mu$ is a renormalization point and $\mbox{\boldmath$\alpha$}$ is the renormalized gauge coupling. For the $\mbox{MS}$ scheme the renormalization constants include only $\varepsilon$-poles in the case $\Lambda=\mu$. The $\overline{\mbox{MS}}$-scheme \cite{Bardeen:1978yd} is obtained by redefining the renormalization point

\begin{equation}
\mu \to \frac{\mu\,\exp(\gamma/2)}{\sqrt{4\pi}},
\end{equation}

\noindent
where $\gamma\equiv - \Gamma'(1)\approx 0.577$. For $\Lambda\ne \mu$ minimal subtraction is obtained if the renormalization constants include only $\varepsilon$-poles and powers of $\ln\Lambda/\mu$. Some multiloop calculations with this regularization can be found in \cite{Aleshin:2015qqc,Aleshin:2016rrr,Aleshin:2019yqj}

To define a field renormalization constant $\mbox{\boldmath$Z$}(\mbox{\boldmath$\alpha$},\varepsilon^{-1})$, we require the finiteness of the corresponding renormalized Green's function $G_R$ in the limit $\varepsilon\to 0$,

\begin{equation}
G_R\Big(\mbox{\boldmath$\alpha$},\ln\frac{\mu}{P}\Big) = \lim\limits_{\varepsilon\to 0}\mbox{\boldmath$Z$}(\mbox{\boldmath$\alpha$},\varepsilon^{-1})\, G\Big[\Big(\frac{\mu}{P}\Big)^\varepsilon\mbox{\boldmath$\alpha$} \mbox{\boldmath$Z_\alpha$}^{-1}(\mbox{\boldmath$\alpha$},\varepsilon^{-1}),\varepsilon^{-1}\Big].
\end{equation}

In $D$-dimensions RGFs are defined according to the prescription

\begin{eqnarray}
&& \mbox{\boldmath$\beta$}(\mbox{\boldmath$\alpha$},\varepsilon) \equiv \frac{d\mbox{\boldmath$\alpha$}(\alpha_0(\Lambda/\mu)^\varepsilon,\varepsilon^{-1})}{d\ln\mu}\bigg|_{\alpha_0=\text{const}};\qquad\nonumber\\
&& \mbox{\boldmath$\gamma$}(\mbox{\boldmath$\alpha$}) \equiv \frac{d\ln \mbox{\boldmath$Z$}(\mbox{\boldmath$\alpha$}, \varepsilon^{-1})}{d\ln\mu} \bigg|_{\alpha_0=\text{const}}.
\end{eqnarray}

\noindent
(In our notations they are denoted in bold.)

Alternatively, it is possible to perform the renormalization in the four-dimensional form

\begin{eqnarray}
&&\hspace*{-5mm} \frac{1}{\alpha_0} = \frac{Z_\alpha (\alpha, \varepsilon^{-1}, \ln \Lambda/\mu)}{\alpha};\nonumber\\
&&\hspace*{-5mm} G_R\Big(\alpha,\ln\frac{\mu}{P}\Big) = \lim\limits_{\varepsilon\to 0} Z(\alpha,\varepsilon^{-1},\ln\Lambda/\mu)\,\nonumber\\
&&\hspace*{-5mm} \qquad\qquad\qquad\qquad \times G\Big[\Big(\frac{\Lambda}{P}\Big)^\varepsilon\alpha Z_\alpha^{-1}(\alpha,\varepsilon^{-1},\ln\Lambda/\mu), \varepsilon^{-1}\Big].\qquad
\end{eqnarray}

\noindent
In this case the renormalization constants $Z_\alpha$ and $Z$ should not contain positive powers of $\varepsilon$, but include powers of $\ln\Lambda/\mu$. In this case RGFs are defined as

\begin{eqnarray}
&& \beta(\alpha) \equiv \frac{d \alpha(\alpha_0, \varepsilon^{-1}, \ln \Lambda/\mu) }{d \ln \mu}\bigg|_{\alpha_0=\text{const}};\qquad\nonumber\\
&& \gamma(\alpha) \equiv \frac{d\ln Z(\alpha,\varepsilon^{-1},\ln\Lambda/\mu)}{d\ln\mu}\bigg|_{\alpha_0=\text{const}}
\end{eqnarray}

\noindent
and are related to the $D$-dimensional ones by the equations

\begin{equation}
\mbox{\boldmath$\beta$}(\alpha,\varepsilon) = - \varepsilon\alpha + \beta(\alpha);\qquad \mbox{\boldmath$\gamma$}(\alpha) = \gamma(\alpha).
\end{equation}

It is important that RGFs $\beta(\alpha)$ and $\gamma(\alpha)$ do not depend on both $\varepsilon$ and $\ln\Lambda/\mu$. From this requirement it is possible to relate the coefficients at higher $\varepsilon$-poles and higher powers of $\ln\Lambda/\mu$ to the coefficients of the $\beta$-function and anomalous dimension. For the regularization considered here the corresponding equations are encoded in the all-order exact formulas \cite{Meshcheriakov:2024qwj}

\begin{eqnarray}\label{Exact_Equation1}
&&\hspace*{-7mm} \ln \alpha_0 = \exp\Big(\ln\frac{\Lambda}{\mu}\,\beta(\alpha)\frac{\partial}{\partial\alpha}\Big)\,
\bigg\{-\int\limits_0^\alpha \frac{d\alpha}{\alpha} \frac{\beta(\alpha)}{\beta(\alpha)-\varepsilon\alpha} +\ln\alpha\bigg\};\\
\label{Exact_Equation2}
&&\hspace*{-7mm} \alpha_0^{-S} = \exp\Big(\ln\frac{\Lambda}{\mu}\,\beta(\alpha) \frac{\partial}{\partial\alpha}\Big)\,\alpha^{-S} \exp\bigg\{S\int\limits_0^\alpha\frac{d\alpha}{\alpha}\frac{\beta(\alpha)}{\beta(\alpha)-\varepsilon\alpha}\bigg\};\\
\label{Exact_Equation3}
&&\hspace*{-7mm} \ln Z - \int\limits^\alpha_a d\alpha\,\frac{\gamma(\alpha)}{\beta(\alpha)} = \exp\Big(\ln\frac{\Lambda}{\mu}\, \beta(\alpha) \frac{\partial}{\partial\alpha}\Big)
\nonumber\\
&&\hspace*{-7mm}
\qquad\qquad\qquad\qquad\qquad \times \bigg[\int\limits_0^\alpha d\alpha\, \frac{\gamma(\alpha)}{\beta(\alpha)-\varepsilon\alpha} - \int\limits^\alpha_a d\alpha\,\frac{\gamma(\alpha)}{\beta(\alpha)}\bigg],\qquad
\end{eqnarray}

\noindent
where the constant $a$ in the last equation can be arbitrary.

This form reveals the renormalization group origin of the considered equations. For this purpose, let us first investigate the case of the cutoff type regularizations (like the regularization by higher covariant derivatives  \cite{Slavnov:1971aw,Slavnov:1972sq}), which is obtained by removing $\varepsilon$-poles in the formal limit $\varepsilon \to \infty$. Then, a renormalization prescription analogous to minimal subtraction is the HD+MSL scheme \cite{Kataev:2013eta}, in which the renormalization constants include only powers of $\ln\Lambda/\mu$. Therefore, choosing $\mu'=\Lambda$ and taking into account that $\alpha'(\mu) = \alpha(\Lambda) = \alpha(\alpha_0,\ln\Lambda/\mu = 0)=\alpha_0$ for an arbitrary function $f(\alpha)$ in the HD+MSL scheme we obtain

\begin{equation}
f(\alpha_0) = \exp\Big(\ln\frac{\Lambda}{\mu}\, \beta(\alpha)\frac{\partial}{\partial\alpha}\Big) f(\alpha).
\end{equation}

\noindent
This equation exactly reproduces the expressions (\ref{Exact_Equation1}), (\ref{Exact_Equation2}), and (\ref{Exact_Equation3}) for $\ln Z_\alpha$, $(Z_\alpha)^S$, and $\ln Z$ if

\begin{equation}
f(\alpha_0) = \ln\alpha_0;\qquad f(\alpha_0) = \alpha_0^{-S};\qquad f(\alpha_0) = \smash{\int\limits_a^{\alpha_0}} d\alpha\, \frac{\beta(\alpha)}{\gamma(\alpha)},
\end{equation}

\noindent
respectively.

Next,  let us consider a more complicated case of the dimensional regularization with $\Lambda\ne \mu$. In this version of dimensional regularization the renormalization constants contain not only $\varepsilon$-poles, but also powers of $\ln\Lambda/\mu$ and various mixed terms. In the standard case $\mu=\Lambda$ from the above equations we see that \cite{tHooft:1973mfk}

\begin{equation}
\alpha\exp\bigg\{-\int\limits_0^{\alpha}\frac{d\alpha}{\alpha}\frac{\beta(\alpha)}{\beta(\alpha)-\varepsilon\alpha}\bigg\}\bigg|_{\mu=\Lambda} = \alpha_0 = \alpha Z_\alpha^{-1}(\alpha,\varepsilon^{-1},0).
\end{equation}

\noindent
Using the exponential map, for an arbitrary value of the renormalization point $\mu$ we obtain

\begin{equation}
f(\alpha_0) = \exp\Big(\ln\frac{\Lambda}{\mu}\, \beta(\alpha)\frac{\partial}{\partial\alpha}\Big) f \Bigg(\alpha \exp\bigg\{-\int\limits_0^{\alpha} \frac{d\alpha}{\alpha} \frac{\beta(\alpha)}{\beta(\alpha)-\varepsilon\alpha}\bigg\}\Bigg).
\end{equation}

\noindent
In the particular case $f(\alpha_0)=1/\alpha_0$ this equation gives

\begin{equation}\label{Z_Alpha}
Z_\alpha(\alpha,\varepsilon^{-1},\ln\Lambda/\mu) = \alpha \exp\Big(\ln\frac{\Lambda}{\mu}\, \beta(\alpha) \frac{\partial}{\partial\alpha}\Big) \Big(\alpha^{-1} Z_\alpha(\alpha,\varepsilon^{-1},0)\Big),
\end{equation}

\noindent
where

\begin{equation}
Z_\alpha(\alpha,\varepsilon^{-1},0) = \exp\bigg(\int\limits_0^\alpha\frac{d\alpha}{\alpha} \frac{\beta(\alpha)}{\beta(\alpha)-\varepsilon\alpha}\bigg).
\end{equation}

After some transformations (see \cite{Meshcheriakov:2024qwj} for details) the expression (\ref{Z_Alpha}) can be cast in the form

\begin{eqnarray}
&& Z_\alpha(\alpha,\varepsilon^{-1},\ln\Lambda/\mu) = \exp\Big\{\ln\frac{\Lambda}{\mu}\, \Big[\beta(\alpha) \frac{\partial}{\partial\alpha} - \frac{\beta(\alpha)}{\alpha}\Big]\Big\}  Z_\alpha(\alpha,\varepsilon^{-1},0)\qquad
\nonumber\\
&& = \exp\Big\{\ln\frac{\Lambda}{\mu}\, \Big[\beta(\alpha) \frac{\partial}{\partial\alpha} - \gamma_\alpha(\alpha)\Big]\Big\}  \exp\bigg(\int\limits_0^\alpha\frac{d\alpha}{\alpha} \frac{\beta(\alpha)}{\beta(\alpha)-\varepsilon\alpha}\bigg),\qquad
\end{eqnarray}

\noindent
where we took into account that $\gamma_\alpha(\alpha) \equiv d\ln Z_\alpha/d\ln\mu = \beta(\alpha)/\alpha$.

This expression is a particular case of the general equation for an arbitrary renormalization constant,

\begin{eqnarray}\label{General_Z_Formula}
&& Z(\alpha,\varepsilon^{-1},\ln\Lambda/\mu) = \exp\Big\{\ln\frac{\Lambda}{\mu}\Big(\beta(\alpha)\frac{\partial}{\partial\alpha} - \gamma(\alpha)\Big)\Big\}\, Z(\alpha,\varepsilon^{-1},0)\qquad\nonumber\\
&& = \exp\Big\{\ln\frac{\Lambda}{\mu}\Big(\beta(\alpha)\frac{\partial}{\partial\alpha} - \gamma(\alpha)\Big)\Big\}\,
\exp\Big\{\int\limits_0^\alpha d\alpha\, \frac{\gamma(\alpha)}{\beta(\alpha)-\varepsilon\alpha}\Big\}.\qquad
\end{eqnarray}

\noindent
As a correctness test, in \cite{Meshcheriakov:2024qwj} it was verified that this equation exactly reproduces the five-loop expression for $\ln Z$ presented in \cite{Meshcheriakov:2023fmk}. Here write down only the four-loop expression, because it is essentially smaller,

\begin{eqnarray}
&&\hspace*{-5mm} \ln Z = - \alpha \gamma_1 \Big(\frac{1}{\varepsilon} + \ln\frac{\Lambda}{\mu} \Big) - \frac{\alpha^2}{2} \bigg[\gamma_2 \Big(\frac{1}{\varepsilon}+ 2 \ln\frac{\Lambda}{\mu} \Big)
+ \gamma_1\beta_1 \Big(\frac{1}{\varepsilon} + \ln\frac{\Lambda}{\mu} \Big)^2  \bigg]
\nonumber\\
&&\hspace*{-5mm} - \frac{\alpha^3}{3} \bigg[ \gamma_3 \Big(\frac{1}{\varepsilon} + 3 \ln\frac{\Lambda}{\mu} \Big)
+ \gamma_1\beta_2 \Big(\frac{1}{\varepsilon^2} + \frac{3}{\varepsilon} \ln\frac{\Lambda}{\mu} + \frac{3}{2} \ln^2\frac{\Lambda}{\mu} \Big)
+ \gamma_2\beta_1 \Big(\frac{1}{\varepsilon^2}
\qquad\nonumber\\
&&\hspace*{-5mm} \qquad + \frac{3}{\varepsilon} \ln\frac{\Lambda}{\mu} + 3\ln^2\frac{\Lambda}{\mu} \Big) + \gamma_1\beta_1^2 \Big(\frac{1}{\varepsilon} + \ln\frac{\Lambda}{\mu} \Big)^3\bigg]
\nonumber\\
&&\hspace*{-5mm}  - \frac{\alpha^4}{4} \bigg[ \gamma_4 \Big(\frac{1}{\varepsilon}+ 4 \ln\frac{\Lambda}{\mu}\Big)
+\gamma_1\beta_3 \Big(\frac{1}{\varepsilon^2} + \frac{4}{\varepsilon} \ln\frac{\Lambda}{\mu} + 2\ln^2\frac{\Lambda}{\mu} \Big)
+\gamma_2\beta_2 \Big(\frac{1}{\varepsilon}
\nonumber\\
&&\hspace*{-5mm} \qquad + 2\ln \frac{\Lambda}{\mu} \Big)^2 + \gamma_3\beta_1 \Big(\frac{1}{\varepsilon^2} + \frac{4}{\varepsilon} \ln\frac{\Lambda}{\mu} + 6\ln^2\frac{\Lambda}{\mu} \Big)
+ 2 \gamma_1\beta_1\beta_2 \Big(\frac{1}{\varepsilon^3} + \frac{4}{\varepsilon^2} \ln\frac{\Lambda}{\mu}
\nonumber\\
&&\hspace*{-5mm} \qquad + \frac{5}{\varepsilon} \ln^2\frac{\Lambda}{\mu} + \frac{5}{3} \ln^3\frac{\Lambda}{\mu} \Big) + \gamma_2\beta_1^2 \Big(\frac{1}{\varepsilon^3} + \frac{4}{\varepsilon^2} \ln\frac{\Lambda}{\mu} + \frac{6}{\varepsilon} \ln^2\frac{\Lambda}{\mu} + 4\ln^3\frac{\Lambda}{\mu} \Big)
\nonumber\\
&&\hspace*{-5mm} \qquad
+ \gamma_1\beta_1^3 \Big(\frac{1}{\varepsilon} + \ln \frac{\Lambda}{\mu} \Big)^4 \bigg] + O(\alpha^5).
\end{eqnarray}

Using Eq. (\ref{General_Z_Formula}) it is possible to compare coefficients at higher poles and logarithms. Namely, according to \cite{Derkachev:2017nhd,Meshcheriakov:2022tyi,Meshcheriakov:2023fmk} the coefficients at higher logarithms for the cutoff type regularizations and at higher $\varepsilon$-poles in the case of using the dimensional technique can be written as

\begin{eqnarray}
&&\hspace*{-3mm} \ln Z_{\alpha}(\alpha,\ln\Lambda/\mu) = - \sum\limits_{q=1}^\infty \sum\limits_{k_1,k_2,\ldots, k_{q}=1}^\infty
\frac{1}{K_{q}}\cdot \frac{K_{q}\mbox{\bf{!}}}{q!}\, \beta_{k_1} \beta_{k_2} \ldots \beta_{k_{q}}\, \alpha^{K_{q}}\, \ln^q \frac{\Lambda}{\mu};\nonumber\\
&&\hspace*{-3mm} \ln \mbox{\boldmath$Z_\alpha$}(\alpha,1/\varepsilon) = - \sum\limits_{q=1}^\infty\, \sum\limits_{k_1,k_2,\ldots, k_{q}=1}^\infty \frac{1}{K_{q}}\, \beta_{k_1} \beta_{k_2} \ldots \beta_{k_{q}}\, \alpha^{K_{q}}\, \varepsilon^{-q},
\end{eqnarray}

\noindent
where $K_m \equiv \sum\limits_{i=1}^m k_i$;\ \ \ $K_m\mbox{\bf{!}}\equiv K_1 K_2 \ldots K_{m}$;\ \ \ $K_0\mbox{\bf{!}}\equiv 1$.

Moreover, it is possible to find some other features of the renormalization constant structure. The simplest one is the relation between coefficients at simple $\varepsilon$-poles and at the first power of $\ln\Lambda/\mu$ given by Eq. (\ref{Simple_Poles}). The coefficients at higher poles and logarithms are related in a much more complicated way, but some features can nevertheless be noted \cite{Meshcheriakov:2023fmk}. As an example, here we consider the expression for $\ln Z_\alpha$, in which all terms proportional to $1/\varepsilon^2$, $\varepsilon^{-1} \ln\Lambda/\mu$, and $\ln^2\Lambda/\mu$ are factorized into perfect squares,

\begin{eqnarray}
&&\hspace*{-4mm} \ln Z_\alpha = -\sum\limits_{L=1}^\infty \alpha^L \beta_L \Big(\frac{1}{L\varepsilon} + \ln\frac{\Lambda}{\mu}\Big) - \frac{1}{L}\sum\limits_{L=2}^\infty \alpha^L \sum\limits_{k=1}^{L-1} \beta_k \beta_{L-k}\Big(\frac{1}{\varepsilon} + \frac{L}{2}\ln\frac{\Lambda}{\mu}\Big)^2
\nonumber\\
&&\hspace*{-4mm} +\ \mbox{higher poles and logarithms}.
\end{eqnarray}

\noindent
It can also be noted that some terms have a rather simple structure

\begin{equation}
\ln Z_\alpha = - \sum\limits_{m=1}^\infty \sum\limits_{k=1}^\infty \frac{\big(\beta_k \alpha^{k}\big)^m} mk \Big(\frac{1}{\varepsilon} + k\ln\frac{\Lambda}{\mu}\Big)^m + \mbox{the other terms}.
\end{equation}

\noindent
Some features have also been found for $Z_\alpha$, $(Z_\alpha)^S$, and $\ln Z$, for instance,

\begin{eqnarray}
&&\hspace*{-5mm} \ln Z = - \sum\limits_{L=1}^\infty \frac{\alpha^{L}}{L}\, \sum\limits_{k=1}^{L}\, \gamma_{L-k+1} (\beta_1)^{k-1} \varepsilon^{L-k}\, \Big(\frac{1}{\varepsilon} + \ln\frac{\Lambda}{\mu}\Big)^L\bigg|_{\varepsilon^s\to 0\ \text{for all}\ s>0} \nonumber\\
&&\hspace*{-5mm} +\ \mbox{terms containing $\beta_i$ with $i\geq 2$}.
\end{eqnarray}

Thus, it is possible to construct simple all-loop equations for the renormalization constants and, using them, analyse the coefficients at higher poles and logarithms.

\section{Conclusion}

Investigation of various renormalization prescriptions is very important, because some exact relations (e.g., the NSVZ equation) are satisfied only in certain subtraction schemes. Different subtraction schemes can be related by finite renormalizations which form a Lie group. The infinitesimal finite renormalizations of the gauge coupling constant form a certain subalgebra of the Witt algebra (the central extension of which is the Virasoro algebra widely used in string theory). It is also possible to construct the commutation relations for the algebra of the infinitesimal finite renormalizations which involve the matter field renormalizations. The non-infinitesimal finite renormalizations can be obtained in standard way with the help of the exponential map. An important particular case of finite renormalizations is the ones that shift the renormalization point $\mu$. They do not change RGFs and can be used for constructing simple formulas for the renormalization constants. This has been done for a version of dimensional regularization in which the dimensionful regularization parameter $\Lambda$ is different from the renormalization point $\mu$. The main advantage of this technique is that it has features of both usual dimensional regularization and the regularizations of the cutoff type. In this case the formulas for the renormalization constants relate the coefficients at $\varepsilon$-poles, powers of $\ln\Lambda/\mu$, and the mixed terms to the coefficients of RGFs, i.e. of the $\beta$-function and anomalous dimension. After setting $\Lambda=\mu$ they reproduce the corresponding results for the usual dimensional regularization, and after removing $\varepsilon$-poles give the expressions for the cutoff type regularizations. They also allow establishing the relation between the coefficients at higher poles and at higher logarithms in a simple way, although at the first sight this relation is highly nontrivial. Using the general equations for the renormalization constants we present the corresponding expressions in the lowest approximations and discuss some general features of the results.


\end{document}